\documentclass[fleqn,usenatbib]{mnras}

\usepackage{newtxtext,newtxmath}

\usepackage[T1]{fontenc}

\DeclareRobustCommand{\VAN}[3]{#2}
\let\VANthebibliography\thebibliography
\def\thebibliography{\DeclareRobustCommand{\VAN}[3]{##3}\VANthebibliography}

\usepackage{graphicx}	
\usepackage{amsmath}	
\usepackage{soul}

\newcommand{\pd}[2]{\frac{\partial #1}{\partial #2}} 
\newcommand{\thd}[3]{\left(\frac{\partial #1}{\partial #2}\right)_{#3}} 
\newcommand{\td}[2]{\frac{d #1}{d #2}}

\title[PNS Convection and NDW r-Processing]{Proto-Neutron Star Convection and the Neutrino-Driven Wind: Implications for the r-Process}

\author[B. Nevins and L. Roberts]{
Brian Nevins,$^{1,2,3}$\thanks{E-mail: nevinsb1@msu.edu}
Luke~F. Roberts$^{4}$
\\
$^{1}$Department of Physics and Astronomy, Michigan State University, East Lansing,
MI 48824, USA\\
$^{2}$Facility for Rare Isotope Beams, Michigan State University, East Lansing, MI
48824, USA\\
$^{3}$Joint Institute for Nuclear Astrophysics – Center for the Evolution of the Elements (JINA-CEE), USA\\
$^{4}$Computer, Computational, and Statistical Sciences Division, Los Alamos National Laboratory, Los Alamos, NM, 87545, USA}

\date{Accepted XXX. Received YYY; in original form ZZZ}

\pubyear{2022}

\begin{document}
\label{firstpage}
\pagerange{\pageref{firstpage}--\pageref{lastpage}}
\maketitle

\begin{abstract}
The neutrino-driven wind from proto-neutron stars is a proposed site for r-process nucleosynthesis, although most previous work has found that a wind heated only by neutrinos cannot produce the third r-process peak. However, several groups have noted that introducing a secondary heating source within the wind can change the hydrodynamic conditions suﬃciently for a strong r-process to proceed. One possible secondary heating source is gravito-acoustic waves, generated by convection inside the proto-neutron star. As these waves propagate into the wind, they can both accelerate the wind and shock and deposit energy into the wind. Additionally, the acceleration of the wind by these waves can reduce the total number of neutrino captures and thereby reduce the ﬁnal electron fraction of the wind. In neutron rich conditions, all of these eﬀects can make conditions more favorable for r-process nucleosynthesis. Here, we present a systematic investigation of the impact of these convection-generated gravito-acoustic waves within the wind on potential nucleosynthesis. We ﬁnd that wave eﬀects in the wind can generate conditions favorable for a strong r-process, even when the energy ﬂux in the waves is a factor of $10^{-4}$ smaller than the total neutrino energy ﬂux and the wind is marginally neutron-rich. Nevertheless, this depends strongly on the radius at which the waves become non-linear and form shocks. We also ﬁnd that both entropy production after shock formation and the acceleration of the wind due to stresses produced by the waves prior to shock formation impact the structure and nucleosynthesis of these winds.
\end{abstract}




\section{Introduction}

After its formation in a core-collapse supernova, a proto-neutron star (PNS) cools by emitting some $10^{53}$ erg in neutrino radiation \citep[see][for a recent review on supernova theory]{Burrows_2021}. After a successful supernova explosion, the PNS is left surrounded by a relatively low density region and it continues to emit neutrinos. Neutrino interactions deposit sufficient energy in the outer layers of the PNS to unbind some material in a neutrino-driven wind (NDW), first predicted by \citet{Duncan_1986}. The NDW is powered mainly by charged-current neutrino interactions, which can both heat the material and alter its neutron-to-proton fraction. Depending on the spectrum of the neutrinos emitted by the cooling PNS, the neutrino-driven wind could become either neutron- or proton-rich. If the wind becomes neutron-rich, there is a possibility for the rapidly outflowing gas to undergo r-process nucleosynthesis (e.g. \citet{Woosley_1994}, \citet{Thompson_2001}, \citet{Wanajo_2013}, and see \citet{Arcones_2012} for a recent review).

The material ejected from the PNS begins in nuclear statistical equilibrium, and as it cools begins forming large numbers of $\alpha$ particles. In a neutron-rich wind, nearly all the protons will be bound into $\alpha$ particles when the wind reaches a temperature of $\approx 5$ GK. Triple-$\alpha$ reactions and subsequent $\alpha$ captures then form a number of heavy 'seed' nuclei, before charged-particle reactions freeze out as the temperature in the wind continues to drop \citep{Woosley_1992}. Free neutrons can then capture onto these seed nuclei as the wind moves outward and produce r-process nuclei \citep{Meyer_1992}. The determining factor for whether a strong r-process can take place is the ratio of free neutrons to seed nuclei when seed formation ends. The three primary factors influencing this are the electron fraction, which sets the free neutron abundance; the entropy of the wind during seed formation; and the dynamical timescale of the wind during seed formation \citep{Hoffman_1997}. The influence of the electron fraction is clear: a strong r-process requires an abundance of free neutrons, and specifically, a high neutron-to-seed ratio so that the heaviest elements can be formed. At constant temperature, a higher entropy implies a lower density. The triple-$\alpha$ and neutron-catalyzed triple-$\alpha$ reactions that form the initial seed nuclei are 3- and 4-body interactions, which scale strongly with density. Thus, a high entropy means that these reactions will be much less efficient, resulting in fewer seeds being formed, and increasing the neutron-to-seed ratio in the wind. Finally, the dynamical timescale of the wind determines how long seed formation can proceed before charged particle reactions freeze out. A short dynamical timescale means that fewer seeds will have the chance to form. A sufficiently short dynamical timescale can also compensate for a lower entropy in this way, and allow an r-process to proceed. The cube of the entropy, divided by the dynamical timescale, has often been used as a criterion for determining r-process feasibility \citep{Qian_1996, Hoffman_1997}. 

The NDW was initially predicted to be neutron rich, and \citet{Woosley_1994} found it underwent a strong r-process that closely matched the solar r-process abundance pattern, in large part due to the high entropies found in their calculations. Subsequent work \citep[e.g.][]{Witti_1994, Qian_1996, Otsuki_2000, Thompson_2001} failed to reproduce conditions suitable for a strong r-process, finding entropies significantly lower than \citet{Woosley_1994}. Later work has explored the impact of other possible physics on the NDW, but has generally shown that, outside of extreme conditions -- high PNS mass, unrealistically low electron fractions, or magnetar-strength magnetic fields -- a wind heated purely by neutrinos does not reach high enough entropies or short enough dynamical timescales during seed formation to allow for a strong r-process \citep[e.g.][]{Thompson_2001, Metzger_2007, Wanajo_2013}. The inclusion of corrections from general relativity tends to make conditions more favorable for the r-process, but a very high PNS mass is still required for a strong r-process to proceed \citep{Cardall_1997, Otsuki_2000, Thompson_2001}. A number of studies have explored the effects of rotation and magnetic fields in varying dimensionality \citep{Metzger_2007,Vlasov_2014,Thompson_2018,Desai_2022}, further confirming that extreme conditions - high PNS masses and magnetar-strength magnetic fields - are required for conditions to favor an r-process.

Other studies have focused on the electron fraction in the wind, as \citet{Hoffman_1997} predicts that a lower $Y_e$ will allow for strong r-processing with lower entropies. The electron fraction is set by the neutrino physics at work in the wind, which has been studied in increasing detail. Simulations by \citet{Fischer_2010} and \citet{Hudepohl_2010} found that the neutrino spectrum from the PNS was likely to result in a proton-rich wind, precluding an r-process altogether. Subsequent work by \citet{Roberts_2012b} and \citet{Martinez_Pinedo_2012} found slightly neutron-rich conditions when nuclear mean field effects were included. Later studies from \citet{Pllumbi_2015} and \citet{Xiong_2019} included neutrino oscillation effects, again finding that only proton-rich or slightly neutron-rich conditions were likely to occur in the wind. In short, it seems unlikely that the generally low seed-formation entropy can be compensated by an increased neutron fraction. 

Rather, the most promising avenue for a strong r-process in the NDW is to invoke a secondary heating effect that takes place in the seed-forming region of the wind \citep{Qian_1996}. \citet{Suzuki_2005} proposed damped Alfv\'en waves as a source for this heating, finding that waves generated by magnetar-strength magnetic fields could deposit sufficient energy in this region to predict a strong r-process. \citet{Metzger_2007} also suggested that a small amount of additional heating from acoustic waves, deposited in the seed-forming region, could drive a strong r-process independent of magnetorotational effects. More recently, \citet{Gossan_2020} suggested that gravito-acoustic waves generated by PNS convection could have an important effect on the dynamics of the NDW. Most recently, supernova simulations by \citet{Nagakura_2020} and \citet{Nagakura_2021} indicate that such convection is a common and significant feature across a broad range of progenitors, so convection-driven effects in the wind are likely to be important in most supernovae. They find that PNS convection is strongest in the first 1-2 seconds post-bounce, then gradually subsides. Gravito-acoustic wave heating is therefore likely to operate in the early stages of the NDW, when it is most likely to be neutron rich \citep[e.g.][]{Roberts_2012b}. These effects are powered by the gravitational contraction of the PNS, which provides an energy reservoir of some $10^{53}$ erg during contraction and deleptonization \citep{Gossan_2020}. Even a small fraction of this binding energy coupling to the wind via wave emission could have a significant impact. 

In light of this, we present here a systematic parameter study of the effects of convection-driven gravito-acoustic waves on the dynamics and nucleosynthetic behavior of the NDW. These waves are excited by convective motions in the PNS as internal gravity waves, which tunnel through the PNS atmosphere and emerge as acoustic waves in the NDW itself. As they propagate through the wind, these waves provide an additional source of stress, driving a faster outflow. They can also shock, efficiently depositing their energy into the wind and acting as a secondary heat source. Our objective in this paper is to determine the conditions in which a strong r-process can take place when the effects of these waves are included. To this end, we assume a spherically symmetric and slightly neutron-rich wind and investigate the impact of varying the energy contained in the waves reaching the wind region, as well as the frequency of the waves, which impacts the radius of shock formation and their subsequent rate of energy deposition. 

The paper is structured as follows: Section \ref{sec:GAW_OOM} outlines the physics behind the generation of these gravito-acoustic waves, and how they deposit energy into the wind. Section \ref{sec:model} describes the equations used to model the wind, and section \ref{sec:method} describes the computational method we use for running the simulations. In section \ref{sec:results} we present our results. Our results show that r-processing will take place in significant regions of the parameter space, for both fiducial and extreme PNS conditions. 

\section{Gravito-acoustic Waves as a Secondary Heating Source}
\label{sec:GAW_OOM}

\begin{figure}
    \centering
    \includegraphics[scale=.5]{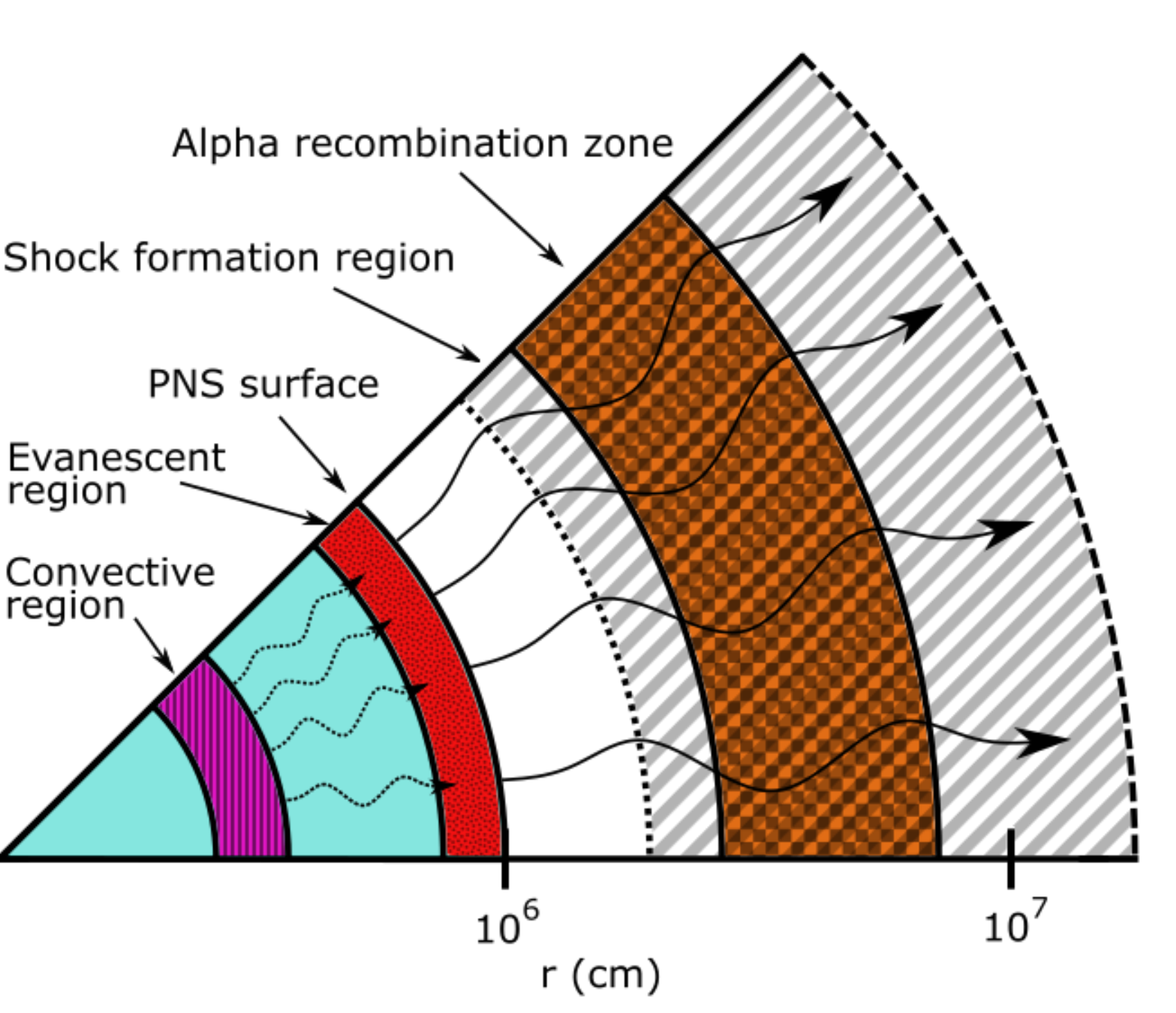}
    \caption{An approximate schematic of gravito-acoustic wave emission and propagation inside and near the PNS. Gravity waves (dashed) are generated by the convective region, attenuate in the evanescent region, and re-emerge as acoustic waves (solid) near the surface of the PNS. They then propagate outward through the wind until they form shocks and dissipate. The region of possible shock formation and heat deposition overlaps with the $\alpha$-forming region. If the waves shock before or during $\alpha$ recombination, the additional heating will inhibit seed formation, making a strong r-process more likely.}
    \label{fig:schematic}
\end{figure}

Shortly after core collapse (${\sim} 200 \, \textrm{ms}$), a convective region develops in the outer mantle of the proto-neutron star \citep{Dessart_2006, Gossan_2020}. Turbulent convection will excite gravito-acoustic waves from the interface between the interior convective region and an exterior radiative region. Both gravity wave modes and acoustic modes will be excited, in addition to non-propagating modes, but due to the Mach number dependence of the wave excitation, the energy flux will be dominated by waves in the gravity wave branch \citep{Goldreich_1990}. The emitted gravity wave luminosity is expected to be 
\begin{equation}
    L_g\approx M_\text{con} L_\text{con}\approx M_\text{con} L_{\nu, \text{tot}}
\end{equation}
where $L_\text{con}$ and $M_\text{con}$ denote the convective luminosity and Mach number, respectively. Furthermore, the convective and total neutrino luminosities $L_\text{con}$ and $L_{\nu,\text{tot}}$ should be approximately equal, as convection is expected to be efficient in the PNS mantle and will carry the majority of the energy flux. $M_\text{con}$ is expected to fall between $10^{-2}$ and $10^{-1}$ \citep{Dessart_2006, Gossan_2020}. Some fraction of the power emitted in gravity waves may propagate from the convective region, through the isothermal atmosphere where the waves will pass through an evanescent region, and into the wind where they will emerge as acoustic waves that can impact the dynamics of the NDW. A schematic of the wave propagation and dissipation in and around the PNS is shown in figure~\ref{fig:schematic}.

Gravity waves are emitted from this convective region with frequencies of $\omega \sim 10^2 - 10^4$ rad s$^{-1}$ \citep{Dessart_2006,Gossan_2020}. 
During the NDW phase of PNS evolution, the convective region is expected to be fairly close to the surface of the PNS \citep{Pons_1999}. The atmosphere of the PNS is nearly isothermal due to neutrino interactions \citep[e.g.][]{Qian_1996} which results in a  Brunt-V{\"a}is{\"a}l{\"a} frequency, $\omega_{BV}$, that is slowly varying with radius up to the point at which the wind is launched. Based on the models described in section \ref{sec:base} and the models of \citet{Roberts_2017}, $\omega_{BV} \sim 5 \times 10^4 \, \textrm{s}^{-1}$. The Lamb frequency in this region is $\omega_L \sim c_s/r \sim 10^3 \, \textrm{s}^{-1}$. Therefore, the waves excited by convection will be evanescent through the PNS atmosphere and emerge into the acoustic branch as the density rapidly falls off in the wind region. 

Using the models of the NDW described in section \ref{sec:base} with no heating and employing the WKB approximation as described in \citet{Gossan_2020}, we find that for a Gaussian distribution of frequencies centered at $10^3\, \textrm{s}^{-1}$, with a standard deviation of $10^2\, \textrm{s}^{-1}$, and angular modes ranging from $\ell=1$ to $\ell=6$ (assuming equal power in each mode), the average transmission efficiency is $\mathcal{T}_\text{avg} = 0.11$. The transmission efficiency ranges between $\mathcal{T}_\text{avg} \sim 0.01 - 0.2$ for a wide range of mean wave frequencies. Rather than try to model this wave transmission in detail, we allow for transmission efficiencies in this range and take $L_w \sim M_\text{con} \mathcal{T} L_{\nu, \text{tot}} \sim 10^{-5} - 10^{-2} L_{\nu, \text{tot}}$. 
Although the work of \citet{Gossan_2020} considered wave propagation in the pre-explosion supernova environment before a NDW had formed, their results for the transmission efficiency are similar to this range of estimates of the transmission efficiency for the post-explosion phase. 
We do not track the evanescent region in our models, but rather assume the waves have an acoustic character throughout the wind. 

The total power of net neutrino heating in the wind is only \citep{Qian_1996}
\begin{equation}
\label{eq:Qnu}
  \frac{\dot Q_\nu}{L_{\nu, \text{tot}}} \sim 1.5 \times 10^{-4} L_{\bar \nu_e, 51}^{2/3} R_6^{2/3} \left(\frac{1.4 M_\odot}{M_\text{NS}}\right),    
\end{equation} 
with $M_\text{NS}$ being the PNS mass, $R_6$ being the PNS radius in units of $10^6 \, \textrm{cm}$, $L_{\bar \nu_e, 51}$ being the electron antineutrino luminosity in units of $10^{51} \, \textrm{erg s}^{-1}$, and assuming an average neutrino energy of 12 MeV. Therefore, based on energetic arguments alone it is clear that the presence of these gravito-acoustic waves is likely to have a significant impact on the dynamics of the wind. There are two ways in which the waves can affect the wind. First, even in the linear regime, the waves will act as a source of stress in the wind \cite[e.g.][]{Jacques_1977} and accelerate the wind. Second, as the waves become non-linear, they will shock and dissipate their energy into heat. By changing the NDW dynamics, both of these effects can alter the nucleosynthetic yields of the wind. A faster outflow reduces the time available for carbon production to occur and will result in a more alpha-rich freeze out \citep{Hoffman_1997}. More heating, if it occurs before alpha recombination, will increase the entropy of the wind and make alpha recombination occur at a lower density, also leading to a more alpha-rich freezeout. An additional source of heat or kinetic energy will also reduce the amount of neutrino capture heating required to unbind the wind material, which will in turn lower the equilibrium electron fraction of the wind. In neutron rich conditions, all of these effects will result in more favorable conditions for r-process nucleosynthesis.

\section{Wind Model}\label{sec:model}

To model the neutrino-driven wind, we solve the equations of general relativistic hydrodynamics in spherical symmetry in steady state. The background metric is assumed to be Schwarzschild sourced by the mass of the PNS $M_\text{NS}$, i.e. we neglect self gravity. These equations are then augmented by a model equation for the evolution of the wave action and its coupling to the background flow, derived following \citet{Jacques_1977}. We seek trans-sonic solutions of the wind equations, so we place the momentum equation in critical form \citep{Thompson_2001}. With these assumptions, the equations of continuity, momentum conservation, entropy ($s$), lepton number conservation, and wave action ($S$) evolution give
\begin{eqnarray}\label{eq:critical_form}
\dot M_\text{NS} &=& 4 \pi r^2 e^\Lambda W v \rho \nonumber\\
 \td{v}{r} &=& \frac{v}{r}\frac{f_2}{f_1} \nonumber\\
 \td{s}{r} &=& \frac{\xi_s}{r} \nonumber\\
 \td{Y_e}{r} &=& \frac{\xi_{Y_e}}{r} \nonumber\\
 \td{S}{r} &=& -S\left(\frac{2}{r}+\frac{1}{l_d}+\frac{1}{v_g}\td{v_g}{r}\right)
\end{eqnarray}
where
\begin{align}\label{eq:dimensional_quantities}
&f_1 = \left(1 - \frac{v^2}{c_s^2} \right) +\delta f_1 \nonumber\\
&f_2 = - \frac{2}{W^2} + \frac{G M_\text{NS}}{c_s^2 r} \frac{1-\left(\frac{c_s}{c}\right)^2}{e^{2\Lambda} W^2} 
 \nonumber\\&\hspace{20pt} +  \frac{1}{W^2 h \rho c_s^2} \left[\xi_s\thd{P}{s}{\rho,Y_e}+\xi_{Y_e}\thd{P}{Y_e}{\rho,s}\right] + \delta f_2\nonumber\\
&\xi_{s} = \frac{r}{v} \frac{\dot{q}_\textrm{tot}}{e^\Lambda W T} \nonumber\\
&\xi_{Y_e} = \frac{r}{v} \frac{\dot{Y}_e}{e^\Lambda W}.
\end{align}
The wave action $S$ is connected to the wave luminosity $L_w$ via
\begin{equation}\label{eq:wave_lum}
    S = \frac{L_w}{4\pi r^2 c_{s} \omega}.
\end{equation}
Here $r$ is the distance of the from the centre of the PNS, and $v$ is the radial velocity of the wind. The total heating rate per baryon is $\dot{q}_\textrm{tot} = \dot{q}_\nu + \dot{q}_w$, where the first term is due to neutrino heating and cooling \citep[see][]{Qian_1996} while the second term is due to wave damping. $\dot{Y}_e$ is the rate of change in the electron fraction of the wind due to neutrino reactions \citep[see][]{Qian_1996}. The wave damping length and frequency are $l_d$ and $\omega$ ($\dot{q}_w$, $l_d$ and $\omega$ are discussed in section \ref{sec:heating} below). $T$, $c_s$, $\rho$, and $h$ denote the local temperature, sound speed, density, and enthalpy, respectively. $W$ is the Lorentz factor, and $v_g = v + c_s$ represents the group velocity of the waves. $G$ and $c$ represent the gravitational constant and the speed of light, and $e^\Lambda=\sqrt{1-\frac{2GM_{NS}}{r c^2}}$. Corrections from the wave stress are denoted by $\delta f_1$ and $\delta f_2$ (see section \ref{sec:stress} below). Without the wave action terms, this system is the same as that of \citet{Thompson_2001}. We employ the equation of state of \citet{Timmes_2000}, which assumes the wind is made up of free protons, neutrons, electrons, positrons, and thermal photons. We search for solutions of these equations that pass through the critical or transonic point where $f_1$ and $f_2$ pass through zero at the the same radius.

\subsection{Wave Stress}\label{sec:stress}
Even in the absence of damping, waves in a stellar atmosphere still exert a force on the medium through which they move. This effect is calculated using the wave action, and adds an extra stress to the momentum equation \citep[e.g.][]{Jacques_1977, Suzuki_2005}. For simplicity, we derive these corrections in the non-relativistic limit.

In the absence of wave stress, the non-relativistic momentum equation for this system is 
\begin{equation}
    v \pd{v}{r} = -\frac{1}{\rho}\pd{P}{r}- \frac{GM_\text{NS}}{r^2}
\end{equation}
Combining this with the other conservation equations yields the non-relativistic critical form equation
\begin{align}\label{eq:nonrel_critical_form}
    \biggl(v^2 - c_s^2\biggl) \td{v}{r} = &\frac{v}{r}\biggl(2c_s^2 - \frac{GM_\text{NS}}{r}\biggr) \nonumber\\ 
&- \frac{v}{\rho}\left[\biggl(\pd{P}{s}\biggr)_{\rho,Y_e} \frac{ds}{dr} +\thd{P}{Y_e}{\rho,s}\frac{dY_e}{dr}\right]
\end{align}
from which we extract the non-relativistic forms of $f_1$ and $f_2$:
\begin{eqnarray}\label{eq:nonrel_fs}
f_1 &=& 1 - \left(\frac{v}{c_s}\right)^2 \nonumber\\
f_2 &=& \frac{GM_\text{NS}}{c_s^2 r} - 2 \nonumber\\
&&+ \frac{r}{\rho c_s^2 }\left[\biggl(\pd{P}{s}\biggr)_{\rho,Y_e} \frac{ds}{dr}+\thd{P}{Y_e}{\rho,s}\frac{dY_e}{dr}\right]
\end{eqnarray}  

The non-relativistic momentum equation including corrections from wave propagation is \citep{Jacques_1977}
\begin{equation}
    \rho v \frac{dv}{dr}+\frac{d}{dr}\left(P+a_1 \mathcal{E}\right)+\frac{\mathcal{E}}{A}\frac{dA}{dr}+\rho \frac{GM_\text{NS}}{r^2}=0 
\end{equation}
where $\mathcal{E}=\frac{c_s}{v_g} \omega S$ is the energy density of the waves and $A=4\pi r^2$. Combined with the other conservation equations, this yields a revised version of the critical form equation:
\begin{align}
    &\left[v^2-c_s^2+\frac{a_1 \mathcal{E} v}{\rho v_g}\left(A_\rho \frac{c_v}{v}-2\right)\right]\td{v}{r}
    \nonumber\\ 
    &=\frac{v}{r}\left(2c_s^2-\frac{GM_\text{NS}}{r}\right)-\frac{v}{\rho}\left[\thd{P}{s}{\rho,Y_e}\td{s}{r}+\thd{P}{Y_e}{\rho,s}\frac{dY_e}{dr}\right]
    \nonumber\\
    &\hspace{10pt}+\frac{a_1 \mathcal{E}v}{\rho r}\left[-2A_\rho\frac{c_v}{v_g}+X_E A_s\frac{c_v}{v_g}+2-\frac{2}{a_1}+\frac{r}{l_d}\right]
\end{align}
with $a_1 = \frac{1}{2}(\gamma+1)$, $c_v=c_s-v$, $X_E=\frac{r}{s}\xi_s$, $A_\rho=\thd{\ln c_s}{\ln \rho}{s}$, and $A_s=\thd{\ln c_s}{\ln s}{\rho}$. In the wave action terms, we have assumed a constant adiabatic index $\gamma$. Note that all terms from equation (\ref{eq:nonrel_critical_form}) are present, with an additional correction term on each side. This allows us to define corrections to the original $f_1$ and $f_2$ functions in equation (\ref{eq:nonrel_fs}):
\begin{align}
    \delta f_1 &= \frac{a_1\mathcal{E}}{\rho c_s^2}\left(2\frac{v}{v_g}-A_\rho \frac{c_v}{v_g}\right)\nonumber\\
    \delta f_2 &= -\frac{a_1\mathcal{E}}{\rho c_s^2}\left[\left(A_s\chi_e-2A_\rho\right)\frac{c_v}{v_g}+2\left(1-\frac{1}{a_1}\right)+\frac{r}{l_d}\right]
\end{align}
These corrections are then applied to the fully relativistic $f_1$ and $f_2$ in equation (\ref{eq:critical_form}). 

\subsection{Wave Heating}\label{sec:heating}
Acoustic waves propagating in the wind can become non-linear and shock heat the wind. We model this shock heating via an effective damping length prescription. Wave heating will only begin when the waves steepen into shocks and begin to dissipate their energy. \citet{Mihalas_1984} provides an integral expression for the radial distance at which this takes place:
\begin{equation}
    \frac{1}{4}(\gamma+1)c_s^{-1}\int_0^r u_0(r')dr' = \frac{\pi c_s}{2\omega}
\end{equation}
where $u_0=\sqrt{\frac{\omega S}{\rho}}$ is the amplitude of the velocity perturbation of the waves and $\gamma$ is the adiabatic index of the background material. Here and elsewhere, $\omega$ represents the angular frequency (in the lab frame) of the waves. We then find the condition for shock formation to be
\begin{equation}\label{eq:shock_condition}
    \int_0^r \sqrt{\frac{\omega S}{\rho}}dr' = \frac{2\pi c_s^2}{\omega(\gamma+1)}.
\end{equation}

In the weak shock limit \citep[e.g.][]{Mihalas_1984}, the energy density of the waves $\epsilon_s$ evolves as
\begin{equation}\label{eq:mihalas_wave_action}
    \nabla\cdot(v_g\epsilon_s)=-\frac{m}{\pi}\omega\epsilon_s
\end{equation}
where $m=(v/c_s)^2-1$ is the reduced Mach number. In a static homogeneous background, the shock can be modeled as a simple saw-tooth wave, with energy density
\begin{equation}
    \epsilon_s = \frac{\gamma P m^2}{3(\gamma+1)^2}.
\end{equation}
In the weak shock limit, we take $\epsilon_s = S/\omega$, which allows us to find an expression for $m$ in terms of local quantities. Combining the wave action evolution in Eqs. (\ref{eq:critical_form}) and (\ref{eq:mihalas_wave_action}), and assuming a constant $\omega$, we find the dissipation length
\begin{equation}\label{eq:ld}
    l_d = \frac{\pi\gamma^2}{\gamma+1}\left(\frac{c_s^2\epsilon}{3\omega^3 S}\right)^{1/2}
\end{equation}
where $\epsilon$ represents the energy density of the wind, excluding rest mass. Once the condition in equation (\ref{eq:shock_condition}) is met, the waves will deposit energy into the wind at a rate
\begin{equation}
    \dot{q}_w = \frac{L_w}{4\pi r^2 \rho l_d} = \frac{c_s}{\rho l_d}\omega S.
\end{equation}

\subsection{Reverse Shock}
As the wind expands outward, it will eventually collide with slow-moving material behind the primary supernova shock, causing a reverse shock in the wind \citep{Arcones_2007}. The radius at which this happens will depend on the dynamics of the explosion, and we treat it as a free parameter with the value $5\times 10^8$cm. The relativistic Rankine-Hugoniot shock conditions then determine the conditions of the post-shock wind:
\begin{eqnarray}
    v_1 \rho_1 W_1 &=& v_2 \rho_2 W_2 \nonumber\\
    W_1^2 h_1 \rho_1 v_1^2 + P_1 &=& W_2^2 h_2 \rho_2 v_2^2 +P_2 \nonumber\\
    W_1 h_1 &=& W_2 h_2
\end{eqnarray}
with all quantities defined as previously, and the subscripts 1 and 2 denoting pre- and post- shock conditions respectively. We treat the post-shock outflow behavior following \citet{Arcones_2012}: for the first second post-shock, density is held constant, with velocity dropping as $r^2$. After the first second, velocity is held constant with density dropping as $r^2$ for the rest of the outflow.

\section{Computational Method}\label{sec:method}
In order to circumvent the critical point singularity in equation (\ref{eq:critical_form}), we introduce a dimensionless integration variable $\psi$ such that 
\begin{align}\label{eq:dimensionless_wind_equations}
\td{\ln r}{\psi} &= f_1 \nonumber\\
\td{\ln v}{\psi} &= f_2 \nonumber\\
\td{\ln T}{\psi} &= \frac{f_1 r}{T}\left[\thd{T}{s}{\rho}\td{s}{r}+\thd{T}{\rho}{s}\td{\rho}{r}\right]\nonumber\\
\td{\ln Y_e}{\psi} &= f_1 \xi_{Y_e} \nonumber\\
\td{\ln S}{\psi} &= -f_1 r \left[\frac{2}{r}+\frac{1}{l_d}+\frac{1}{v+c_s}\left(\td{v}{r}+\td{c_s}{r}\right)\right] \nonumber\\
\frac{dI}{d\psi} &= \sqrt{\frac{\omega S}{\rho}}f_1 r
\end{align}
with $f_1$ and $f_2$ defined as previously. We recast the entropy evolution into a temperature evolution equation, as our EOS is formulated in terms of the Helmholtz free energy. The integral in equation (\ref{eq:shock_condition}) is converted to a similar form, with $I=\int_0^r \sqrt{\frac{\omega S}{\rho}}dr'$. Because our system of equations is relatively stiff, we use an interpolation function between an infinite dissipation length (i.e. no shock heating) and the physical value in equation (\ref{eq:ld}) to control the activation of shock heating in the wind.

The six differential equations in equation (\ref{eq:dimensionless_wind_equations}) are integrated with respect to $\psi$ using a 4th order SDIRK method \citep{Kennedy_2016}, with the wind dynamics adjusted at the appropriate points for the reverse shock and subsequent outflow. The starting radius is set to be the surface of the neutron star (fixed at $r_0 = 1 \times 10^6 \, \textrm{cm}$), and the starting density is set to be the surface density of the neutron star (fixed at $\rho_0 = 1 \times 10^{12} \, \textrm{g/cm}^3$). We assume the wind begins in heating-cooling equilibrium, which fixes the starting temperature and electron fraction \citep[see][]{Qian_1996}. The starting value for the wave action is determined by the wave luminosity per equation (\ref{eq:wave_lum}), which we treat as a fraction of neutrino luminosity and a free parameter. The initial wind velocity is also treated as a free parameter, and maps directly to the PNS mass loss rate $\dot{M_\text{NS}}= 4 \pi r_0^2 W e^\Lambda v_0 \rho_0$.

In order to find the critical (or transonic) solution of the wind equations, $f_1$ and $f_2$ must pass through zero at the same radius. We use a shooting method to determine the critical mass loss rate (i.e. $v_0$) for a given parameter set via a one-dimensional rootfinder. We map a given $v_0$ to the value of $\max \{f_1, 0\}- \max \{f_2, 0\}$ at the minimum radius for which $f_1$ passes through zero, for the profile that is generated by that specific $v_0$. The root of this function is the critical velocity, for which $f_1$ and $f_2$ pass through zero simultaneously. Once the critical velocity (or mass loss rate) is found, the full integration is run to a maximum radius of 10$^{10}$ cm. The sensitivity of the evolution equations makes it numerically impossible to actually generate the critical solution, as every solution appears as either a breeze solution or an unphysical one that returns to the initial radius, even when $v_0$ is obtained to machine precision. To circumnavigate this issue, we assign $f_1$ and $f_2$ to their absolute values for the full integration. This produces the correct behavior to machine precision for the transonic solution. 

The temperature and density versus time for a Lagrangian observer are then extracted from the resulting steady state wind profile (and extended to late times with a $t^{-3}$ power law). These profiles are then used to perform calculations of nucleosynthesis using the nuclear reaction network code SkyNet \citep{Lippuner_2017}. The reaction network calculations include strong, weak, symmetric fission, and spontaneous fission reactions, with inverse reactions calculated via detailed balance. 

The input parameters for our models are the PNS mass ($M_\text{NS}/M_\odot \in [1.4, 2.1]$), the total neutrino luminosity ($L_\nu \in$ [$3\times 10^{52}, 1.2\times 10^{53}$] erg s$^{-1}$), and the wave luminosity as a fraction of total neutrino luminosity ($L_w/L_\nu \in [10^{-5}, 10^{-2}]$). We also examine the impact of different wave frequencies in the range of $10^2$ to $10^4$ $s^{-1}$.  We assume that the neutrinos have equal luminosities in all flavors, a zero chemical potential Fermi-Dirac spectrum, and the average electron neutrino energy is fixed at 12 MeV as measured at the surface of the PNS. The average electron antineutrino energy is chosen such that the equilibrium electron fraction of the wind, $Y_{e,\textrm{eq}} = \lambda_{\nu_e}/(\lambda_{\nu_e} + \lambda_{\bar \nu_e})$, takes a target value \citep[see][]{Qian_1996}, where $\lambda_{\nu_e}$ and $\lambda_{\bar \nu_e}$ are the electron neutrino and antineutrino capture rates, respectively. We primarily consider $Y_{e, \textrm{eq}}=0.48$, unless otherwise noted. Relativistic corrections are included in the neutrino capture rates as in \citet{Thompson_2001}.

\section{Results}\label{sec:results}
\subsection{Models without Wave Heating}
\label{sec:base}
\begin{figure}
    \centering
    \includegraphics[scale=1]{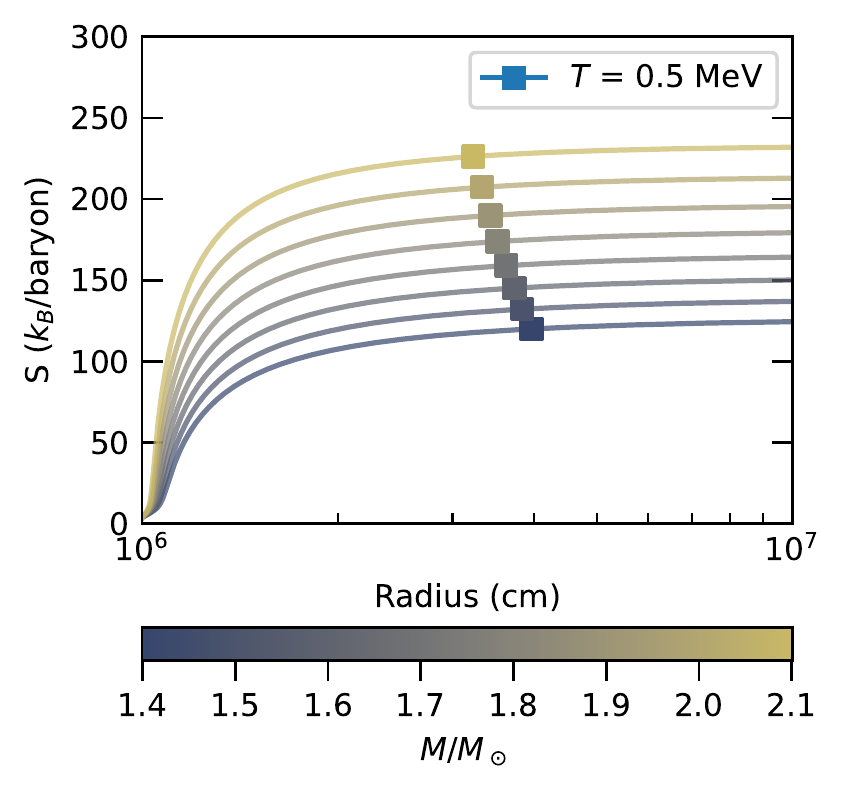}
    \caption{Entropy versus radius for PNSs of varying mass, with a fixed total neutrino luminosity of $6\times 10^{51}$ erg s$^{-1}$, and $L_w = 0$. We find comparable behavior to \citet{Wanajo_2013}. The approximate beginning of seed formation for each model is marked with a square.}
    \label{fig:base_ent}
\end{figure}
\begin{figure}
    \centering
    \includegraphics[scale=1]{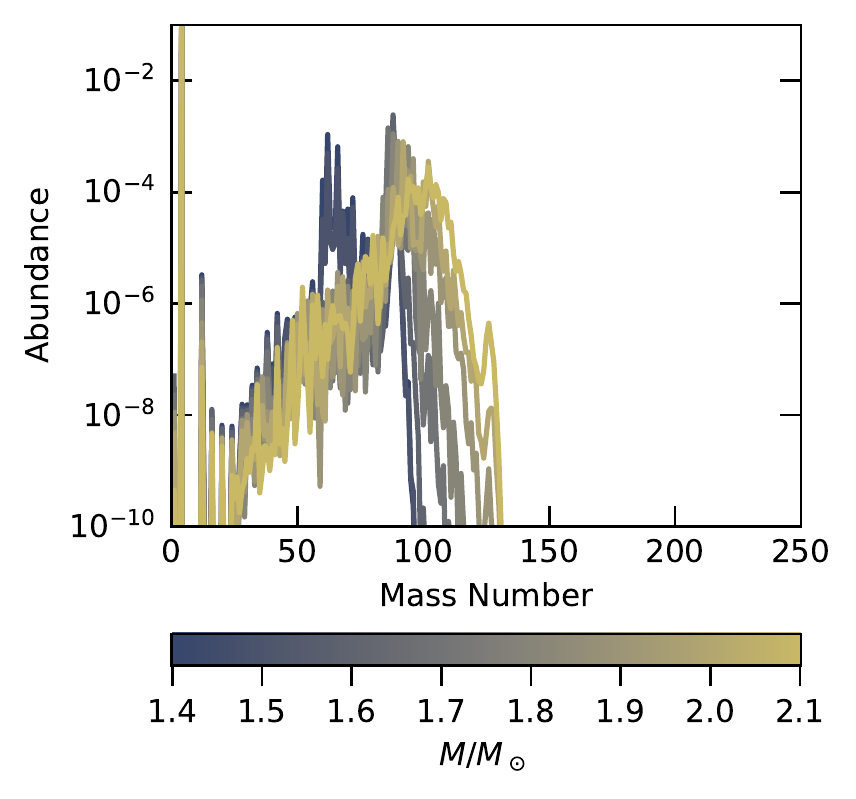}
    \caption{Final abundances in the absence of wave effects (i.e. $L_w=0$), with $L_\nu = 6 \times 10^{51} \, \textrm{erg s}^{-1}$ and PNS masses ranging from 1.4-2.1 $M_\odot$. We find that no r-processing takes place for proto-neutron stars of reasonable masses with $Y_{e, \textrm{eq}} = 0.48$, when wave effects are excluded.}
    \label{fig:no_heat_param_space}
\end{figure}
In the absence of wave contributions (i.e. $L_w = 0$), we find general agreement with prior work regarding the hydrodynamic structure of the wind \citep[e.g.][]{Thompson_2001, Wanajo_2013}. These models serve as a baseline for comparison with the wave heating models shown in subsequent sections. Figure~\ref{fig:base_ent} shows a set of radial entropy profiles for varied PNS masses with a fixed neutrino luminosity of $6 \times 10^{51} \, \textrm{erg s}^{-1}$. Increased PNS mass leads to overall higher entropies throughout the wind \citep{Qian_1996}, which decreases the efficiency of seed formation and brings conditions closer to those required for an r-process. The included general relativistic corrections to the wind equations increase the entropy as expected \citep{Cardall_1997, Thompson_2001}. Nucleosynthesis results for these NDW profiles assuming $Y_{e,\textrm{eq}} = 0.48$ are shown in figure~\ref{fig:no_heat_param_space}. For these models without gravito-acoustic wave heating, the electron fraction at $T=0.5 \, \textrm{MeV}$ is nearly equal to the chosen $Y_{e,\textrm{eq}}$. In contrast to \citet{Wanajo_2013}, we find that even for the highest neutron star masses, no r-processing takes place in these winds.

\subsection{The Impact of Gravito-Acoustic Waves on the NDW}
\subsubsection{Wind Dynamics}
\label{sec:wind_dynamics}
\begin{figure*}
    \centering
    \includegraphics[scale=1]{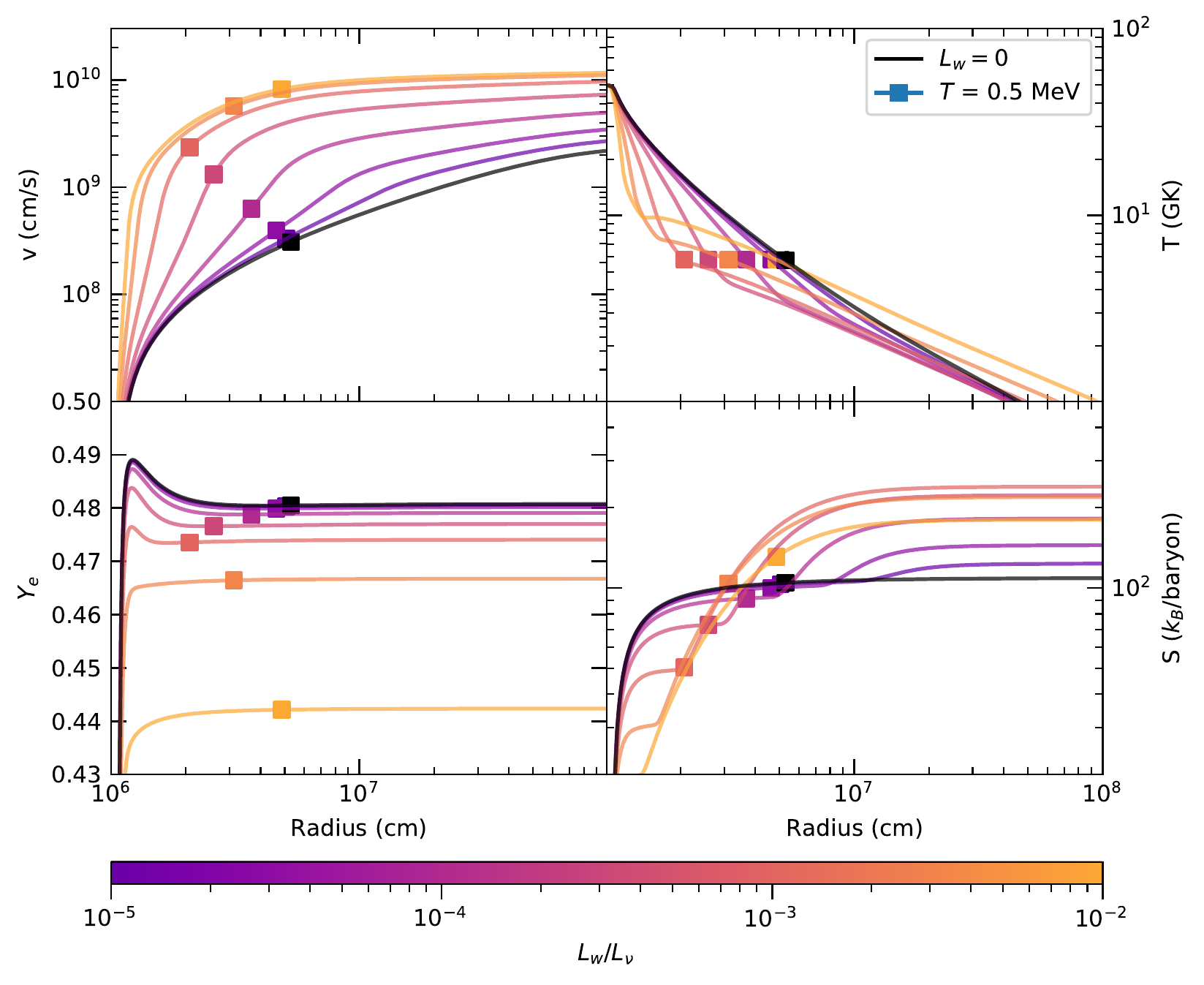}
    \caption{Radial profiles of the velocity, density, temperature, and entropy in the NDW. Different lines correspond to different $L_w$. Other parameters in the wind models were fixed to $M_\text{NS} = 1.5 \, M_\odot$, $L_\nu = 3 \times 10^{52} \, \textrm{erg s}^{-1}$, and $\omega = 2 \times 10^3 \, \textrm{rad s}^{-1}$. The beginning of seed formation for each model is marked a square.}
    \label{fig:fid_profiles}
\end{figure*}

\begin{figure}
    \centering
    \includegraphics[scale=1]{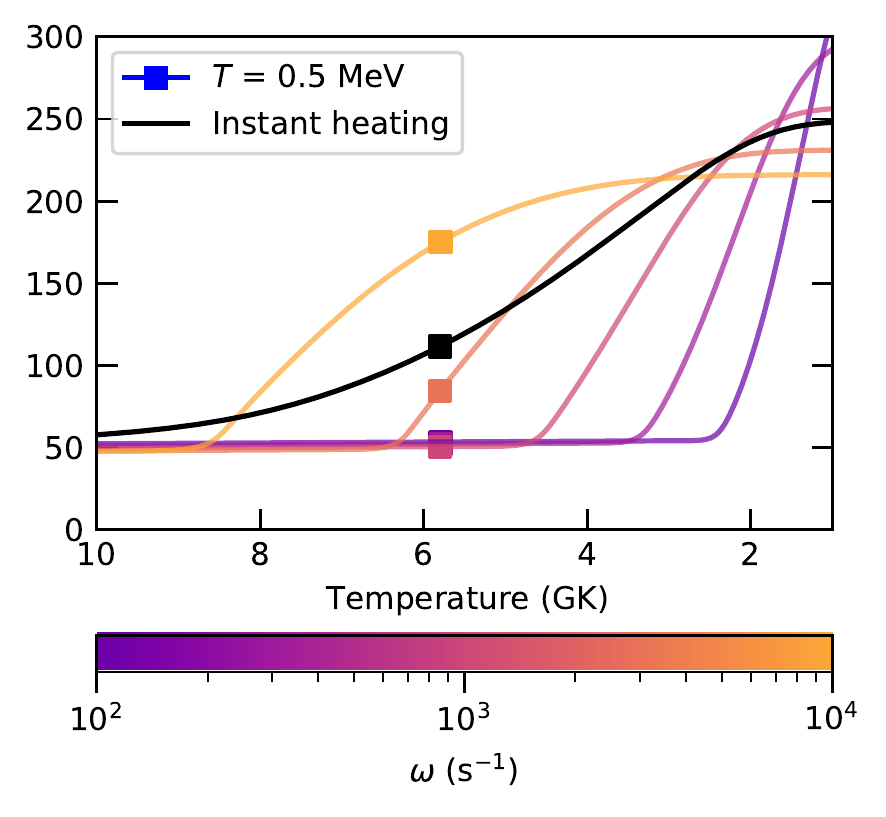}
    \caption{Early entropy profiles for a 1.5 $M_\odot$ neutron star with $L_\nu=3\times 10^{52}$ erg s$^{-1}$ and $L_w=10^{-3}L_\nu$ with varied wave frequencies. For higher frequencies, the shock heating begins to increase the entropy in the wind earlier and has a larger impact where seed nuclei are formed. The impact of the shock prescription is illustrated by the black line, which shows the evolution of the entropy if the waves (with $\omega=2\times 10^3$ rad s$^{-1}$) shock immediately instead of when equation \ref{eq:shock_condition} dictates.}
    \label{fig:freq}
\end{figure}

We now consider the impact of gravito-acoustic waves on the dynamics of the NDW. As is described above, the presence of these waves in the wind can accelerate the NDW by purely mechanical effects and can deposit heat in the wind once the waves shock. Since $L_w$ should scale with $L_\nu$ (see section~\ref{sec:GAW_OOM}), we present our results in terms of the ratio $L_w / L_\nu$. In figure~\ref{fig:fid_profiles}, properties of steady state NDW models with $L_\nu = 3 \times 10^{52} \, \textrm{erg s}^{-1}$, $Y_{e, \textrm{eq}} = 0.48$, $M_\text{NS} = 1.5 \, M_\odot$, $\omega = 2\times 10^3 \, \textrm{rad s}^{-1}$, and varied $L_w/L_\nu$ are shown. Results for $M_\textrm{NS} = 1.9 \, M_\odot$ are qualitatively similar, albeit with higher final entropies. Seed formation begins approximately when the temperature in the wind drops to $T=0.5$ MeV \citep{Qian_1996}, which is marked in figures with a square. Clearly, above $L_w / L_\nu \approx 10^{-5}$, the inclusion of wave effects has a significant impact on the dynamics of the wind. Although $L_w$ in these models is a relatively small fraction of the total neutrino luminosity, it is a large fraction of the neutrino energy that couples to the wind, $\dot Q_\nu$ (see equation~\ref{eq:Qnu}). At small radii, before the waves shock, they accelerate the NDW but do not provide any heating. This results in increasing velocities with $L_w / L_\nu$, and therefore lower densities at a given radius by the relation $\dot M_\text{NS} = 4 \pi r^2 e^{\Lambda} W \rho v$. Additionally, since the acceleration of the wind is no longer provided solely by neutrino heating, the amount of neutrino heating that occurs is lowered, which results in both lower entropies before the wave-heating activation radius, and in lower electron fractions at all points in the wind. As the wave contribution increases, fewer neutrino captures are required to unbind material from the potential well of the PNS and the NDW is accelerated to higher velocities at smaller radii. Both of these effects work to reduce the number of weak interactions in the wind and prevent the electron fraction in the wind from reaching $Y_{e, \textrm{eq}}$, which results in more neutron-rich conditions at the beginning of nucleosynthesis. The changes in $Y_e$ begin prior to the waves forming weak shocks, indicating that the wave stress, rather than shock heating, is the primary contributor. These effects will therefore be present regardless of any uncertainty in the shock heating mechanism. We observe a spike in $Y_e$ at small radii due to electron-positron capture when degeneracy is lifted at high temperatures. The electron fraction then relaxes towards $Y_{e,\textrm{eq}}$, but may not reach it due to the wave contributions. 

Subsequent to the waves shocking, the entropy rapidly increases in all models. Shock formation occurs at temperatures between $2$ and $10 \, \textrm{GK}$ depending on $L_w/L_\nu$ (and $\omega$, see figure~\ref{fig:freq}). The extra entropy production provided by $\dot q_w$ is large compared to neutrino heating because of the low temperatures at which it occurs compared to the temperatures where the bulk of the neutrino heating takes place ($\sim 30 \, \textrm{GK}$ in our simulations).  For the largest $L_w/L_\nu$, the entropy can reach asymptotic values of 300, which is quite large compared to even the largest entropies found for models that do not experience wave heating (see section~\ref{sec:base}). Nevertheless, a significant amount of the entropy production occurs during or after the temperatures over which seed nuclei for the r-process are produced ($\sim$ 2 - 8 GK $\, \textrm{GK}$). Therefore, estimating the likelihood of r-process nucleosynthesis from the often used metric $s^3/\tau_d$ \citep[see][]{Hoffman_1997} is difficult as $s$ is no longer nearly constant while seed production occurs. Before the shock formation radius, the waves reduce both $s$ and $\tau_d$\footnote{We define the dynamical timescale $\tau_d$ at a given point in the wind as $T/\dot{T}$, similar to the $r/\dot{r}$ used by \citet{Hoffman_1997}.}. This can hinder or abet an alpha-rich freezeout depending on the relative strength of these two effects. After shock formation, $s$ is increased relative to the $L_w=0$ case, but potentially at temperatures that are too low to impact the alpha-richness of the NDW. Therefore, to better understand the impact of gravito-acoustic wave heating on the wind, detailed nucleosynthesis calculations are required.

The radius at which the waves shock and the rate at which they damp will depend on their frequency content, with the shock formation radius approximately scaling as $\omega^{-1}$ (see equation~\ref{eq:shock_condition}) and the damping length $l_d \propto \omega^{-1}$ for a fixed $L_w$. Therefore, larger wave frequencies will result in wave heating impacting the thermodynamic conditions of the NDW at smaller radii and higher temperatures. In figure~\ref{fig:freq}, we show the impact of varying $\omega$ on the entropy of the wind. Clearly, larger $\omega$ results in a higher entropy at higher temperature, which is potentially more favorable for an alpha-rich freezeout. The limiting case ($\omega \rightarrow \infty$) corresponds to instantaneous shock formation in the wind, but also implies a damping length that goes to zero. Nevertheless, we also show a case with fixed $\omega$ in $l_d$ but assuming instantaneous shock formation, as this has been assumed in previous work looking at secondary heating mechanisms in the NDW \citep{Suzuki_2005, Metzger_2007}. It is not clear what shock formation radii are favored, given the uncertainty in the range of frequencies excited by PNS convection and the approximate nature of equation~\ref{eq:shock_condition}.

\subsubsection{Nucleosynthesis}\label{sec:nuc}

\begin{figure}
    \centering
    \includegraphics[scale=1]{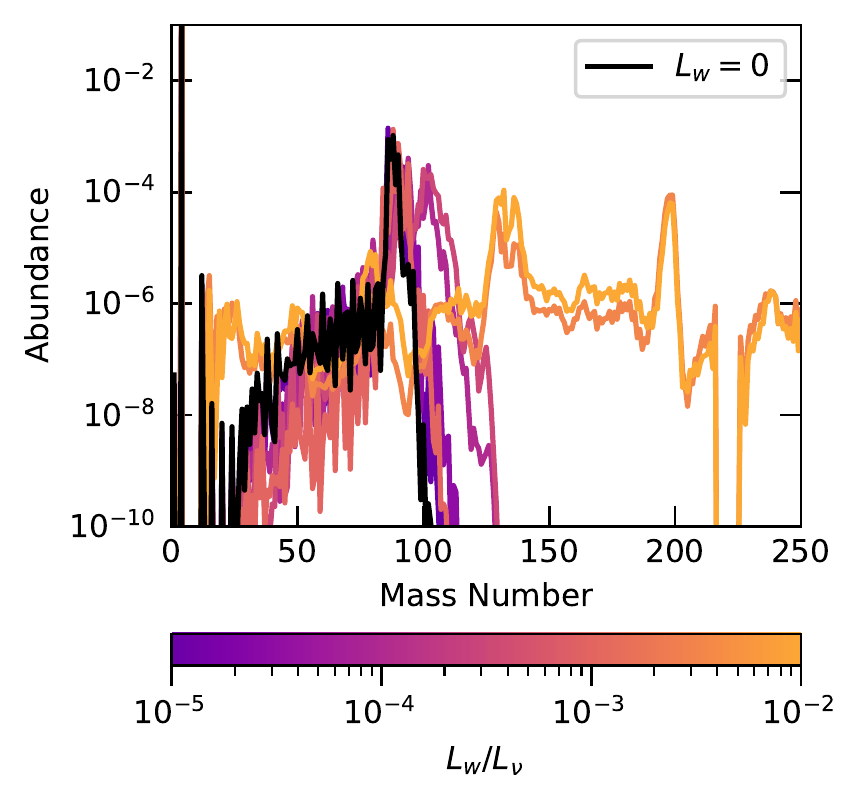}
    \caption{Final nucleosynthesis results, using temperature and density profiles for a 1.5 $M_\odot$ neutron star, with $L_\nu=3\times 10^{52}$ erg s$^{-1}$ and using a wave frequency of $2\times 10^3$ rad s$^{-1}$. A clear peak around mass 200 is indicative of a strong r-process taking place.}
    \label{fig:fid_skynet}
\end{figure}

\begin{figure}
    \centering
    \includegraphics[scale=1]{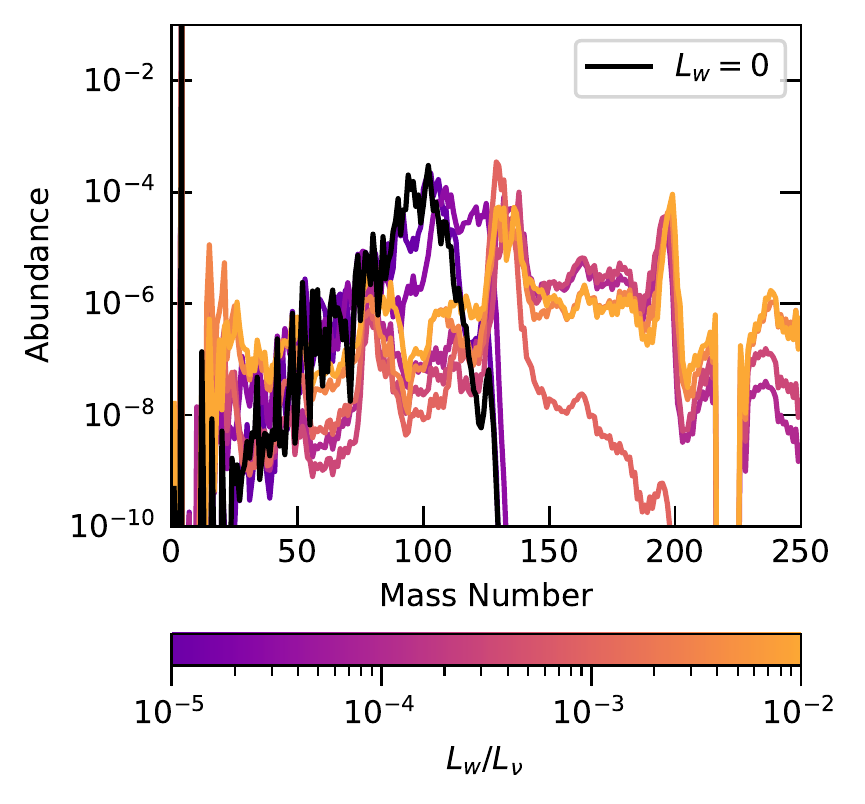}
    \caption{Final nucleosynthesis results, using temperature and density profiles for a 1.9 $M_\odot$ neutron star, with $L_\nu=6\times 10^{52}$ erg s$^{-1}$ and using a wave frequency of $2\times 10^3$ rad s$^{-1}$. A clear peak around mass 200 is indicative of a strong r-process taking place.}
    \label{fig:ext_skynet}
\end{figure}

\begin{figure}
    \centering
    \includegraphics[scale=1]{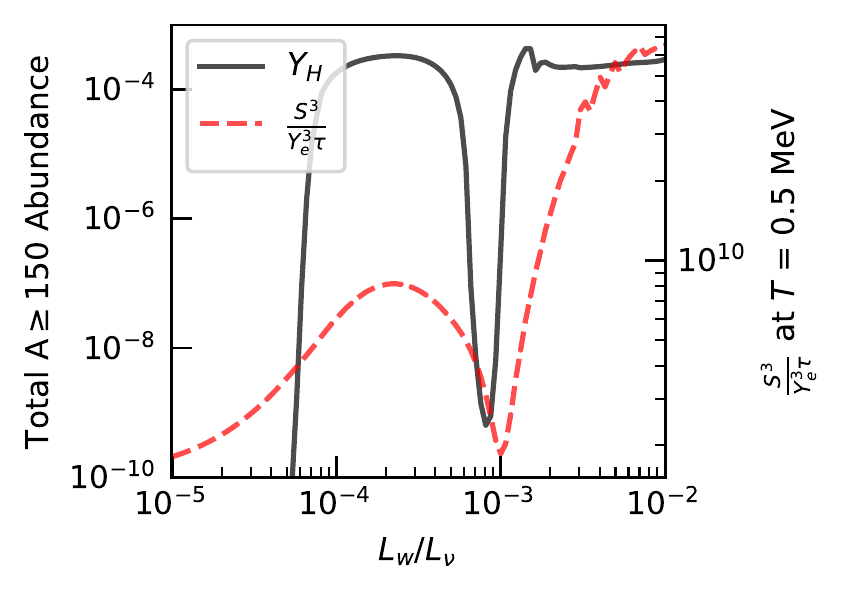}
    \caption{Comparison of the total, summed final abundances of all nuclides with mass $A\geq 150$ (representative of the strength of any r-process taking place) with the quantity $s^3/Y_e^3\tau_d$ evaluated when seed formation begins. The relationship between $s^3/Y_e^3\tau_d$ and $Y_H$ is necessarily approximate due to the presence of wave heating during seed formation. The relationship found in \citet{Hoffman_1997} was derived under the assumption of constant entropy, which is not generally true in our models. Nevertheless, we still observe a strong correlation between the two quantities, which helps to provide a qualitative explanation for the variation in heavy element nucleosynthesis near $L_w/L_\nu=10^{-3}$. These results are for the same parameters as those in figure~\ref{fig:ext_skynet}, with a finer grid in $L_w/L_\nu$.}
    \label{fig:s3tau}
\end{figure}

We now present nucleosynthesis calculations based on the steady-state, gravito-acoustic wave-heated NDW models described in the previous section. Throughout, we assume $Y_{e, \textrm{eq}} = 0.48$ (unless otherwise noted), given that models of neutrino emission from PNSs suggest the NDW will at most be marginally neutron rich. Note that for larger $L_w/L_\nu$ the actual value of $Y_e$ at the beginning of nucleosynthesis can substantially differ from $Y_{e, \textrm{eq}}$ (see figure~\ref{fig:fid_profiles}).

First, we consider the impact of varying $L_w/L_\nu$ for a fixed $\omega = 2 \times 10^3 \, \textrm{rad s}^{-1}$. The final abundances for NDW models with $M_\text{NS}=1.5 M_\odot$ and $L_\nu = 3 \times 10^{52}\text{ erg s}^{-1}$ are shown in figure~\ref{fig:fid_skynet}. These correspond to the NDW models shown in figure~\ref{fig:fid_profiles}. In the absence of wave heating, this parameter set only undergoes an $\alpha$-process that terminates with a peak around mass 90 \citep{Woosley_1992} and is far from the conditions necessary for producing the third r-process peak. Increasing $L_w$, we find that the peak of the abundance distribution increases in mass until $L_w/L_\nu \approx 10^{-4}$. Further increase of $L_w$ from this point briefly reduces the mass of the peak of the abundance distribution, but above $L_w/L_\nu\approx 10^{-3}$ a strong r-process emerges.  The final abundances for NDW models with $M_\text{NS} = 1.9 M_\odot$ and $L_\nu = 6 \times 10^{52}$ are shown in figure~\ref{fig:ext_skynet}. Between $L_w/L_\nu = 10^{-5}$ and $L_w/L_\nu = 10^{-4}$, these models produce both the second and third r-process peaks, but between $L_w/L_\nu \approx 10^{-4}$ and $L_w/L_\nu \approx 10^{-3}$ production of the third peak is again cutoff and the peak of the abundance distribution is pushed down to lower mass. As $L_w/L_\nu$ is increased above $10^{-3}$, a strong r-process re-emerges.

For both sets of parameters, we find the interesting behavior that r-process nucleosynthesis is inhibited for $L_w/L_\nu$ in the approximate range of $10^{-4}$ - $10^{-3}$. This turnover in the maximum mass number is due to the competition between the decreasing dynamical timescale ($\tau_d$) with $L_w$, which inhibits seed formation, and the decreasing entropy ($s$) with $L_w$, which facilitates seed production by increasing the density at which alpha recombination occurs \citep{Hoffman_1997}. Figure~\ref{fig:s3tau} illustrates the correlation between the quantity $s^3/Y_e^3\tau_d$ and the total abundance above mass 150. Despite entropy no longer being constant during seed formation, we do observe a fairly strong correlation between r-process strength and this quantity. We find that as the wave luminosity is increased, $\tau_d$ decreases slightly faster than the entropy, but eventually asymptotes to a minimum value of a few times $10^{-4}$ s. The entropy continues to steadily decrease, which creates the trough in $s^3/Y_e^3\tau_d$ as a function of $L_w$ and gives rise to the window of inhibited r-processing we observe around $L_w/L_\nu=10^{-3}$. At higher $L_w$, shock heating begins prior to alpha recombination, drastically increasing the entropy. This, coupled with the reduced electron fraction at high $L_w$, reinvigorates a strong r-process.

\begin{figure}
    \centering
    \includegraphics[scale=1]{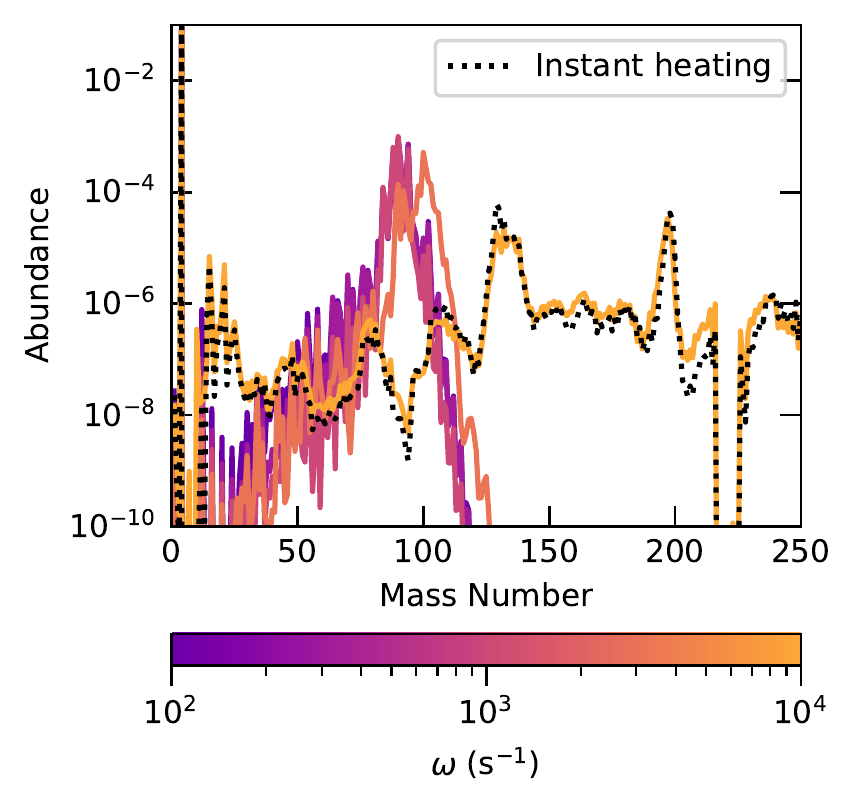}
    \caption{Final abundances for the NDW profiles shown in figure \ref{fig:freq}. For high frequencies, the shock heating begins early enough to drive a strong r-process even for a $1.5 M_\odot$ neutron star. Instantaneous shock formation is illustrated by the black dashed line, showing the final abundances for a wind that immediately experiences shock heating from waves with $\omega = 2\times 10^3$ rad s$^{-1}$.}
    \label{fig:freq_skynet}
\end{figure}

\begin{figure}
    \centering
    \includegraphics[scale=1]{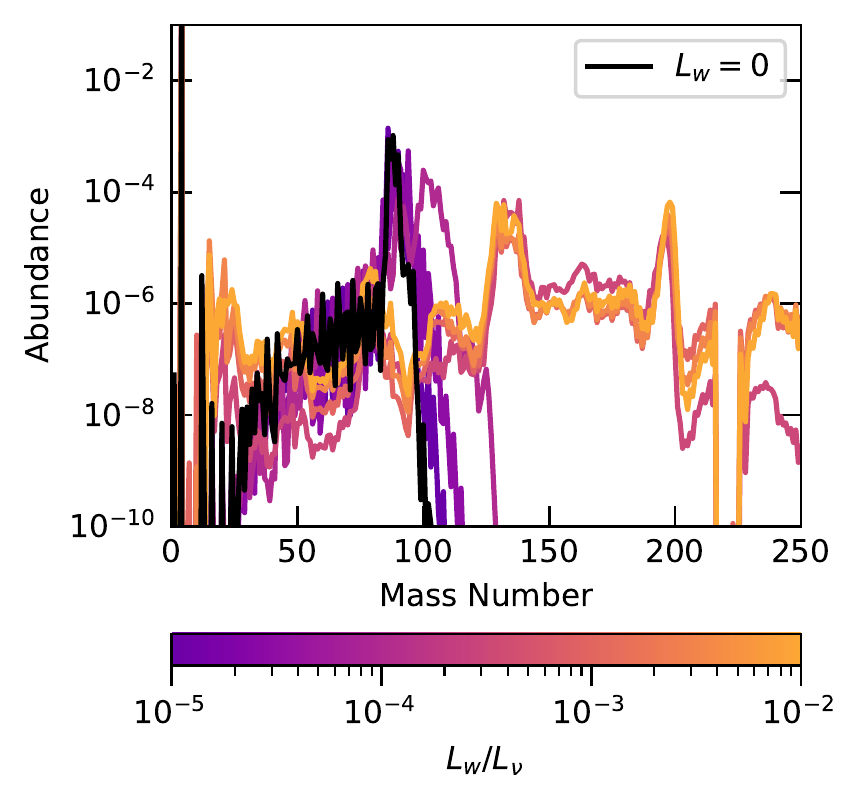}
    \caption{Final abundances using the same parameters as in figure~\ref{fig:fid_skynet}, but assuming that shock heating begins instantaneously in the wind. We see that a strong r-process takes place even for moderate $L_w$.}
    \label{fig:instant_skynet}
\end{figure}

\begin{figure}
    \centering
    \includegraphics[scale=1]{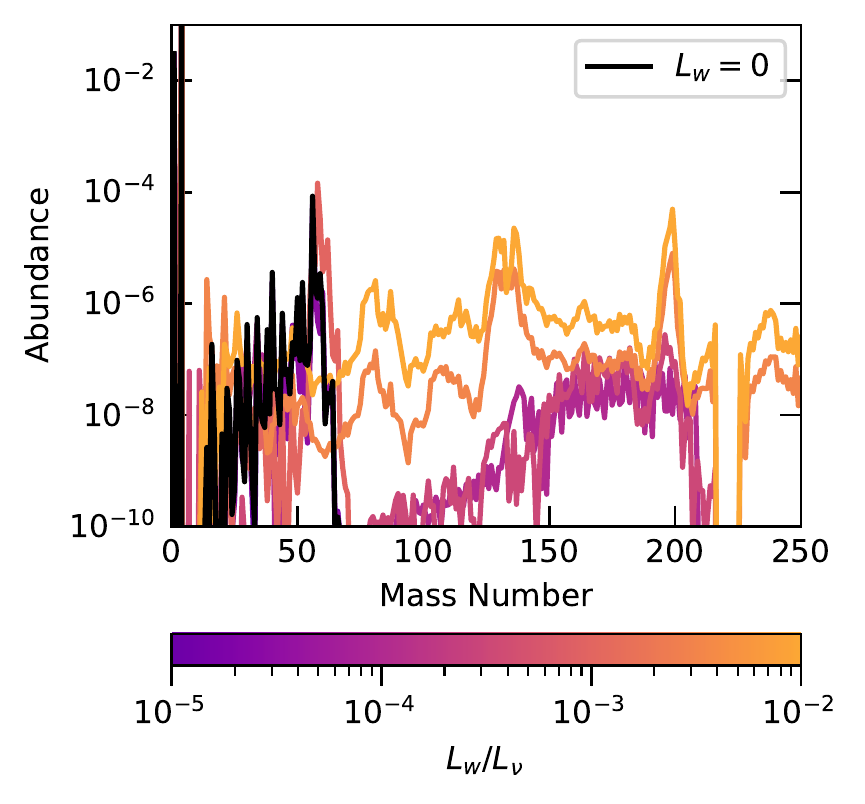}
    \caption{Final abundances using the same parameters as in figure~\ref{fig:ext_skynet}, but with antineutrino energies tuned to $Y_{e, \textrm{eq}}=0.52$. We see r-processing regimes appear, despite a neutrino spectrum that would otherwise have precluded r-processing entirely.}
    \label{fig:ye_52}
\end{figure}
Second, we consider the impact of varying $\omega$ on gravito-acoustic NDW nucleosynthesis. As was noted above, increasing $\omega$ results in an earlier activation of shock heating. In figure~\ref{fig:freq_skynet}, we show the final abundances for $M_\text{NS} = 1.5 \, M_\odot$, $L_\nu = 3 \times 10^{52} \, \textrm{erg s}^{-1}$, and $L_w/L_\nu = 10^{-3}$. For $\omega < 10^4 \, \textrm{s}^{-1}$, the nucleosynthesis is similar to the models with $L_w/L_\nu \approx 10^{-3}$ that efficiently form seed nuclei, as discussed in the preceding paragraphs. Comparing to figure~\ref{fig:freq}, shock heating begins only after the beginning of seed formation and therefore the resulting increase in entropy only has a limited impact on the nucleosynthesis. On the other hand, for the largest frequency considered ($\omega = 10^4 \, \textrm{rad s}^{-1}$), a full r-process pattern extending through the third peak is produced. Here, the wave heating due to weak shocks begins before the start of seed formation. Therefore, the substantial increase in the entropy inhibits seed formation, and leaves a large neutron-to-seed ratio when alpha capture ends. This is mainly driven by the impact of $\omega$ on the shock heating activation radius, and less so by the variation in $l_d$ with $\omega$. This is illustrated by the model shown in figure~\ref{fig:freq_skynet} that assumes $\omega = 2 \times 10^3 \, \textrm{rad s}^{-1}$ but an instantaneous activation of shock heating. This results in nucleosynthesis that is very similar to the $\omega = 10^4 \, \textrm{rad s}^{-1}$ model. 

Therefore, as a limiting case given the uncertainty in the shock activation radius and to compare to previous work \citep{Suzuki_2005, Metzger_2007}, we show in figure~\ref{fig:instant_skynet} final abundances for varied $L_w/L_\nu$ for $\omega = 2 \times 10^3 \, \textrm{rad s}^{-1}$, $M_\text{NS}=1.5 M_\odot$, $L_\nu = 3 \times 10^{52}\text{ erg s}^{-1}$, but with instantaneous activation of the shock heating. The results are noticeably different than those shown in figure~\ref{fig:fid_skynet}, which shows models with the same parameters but without instantaneous shock heating. For instantaneous activation, the average mass of the abundance distribution increases monotonically with $L_w/L_\nu$ and for even moderate wave luminosities is able to produce a full r-process. This illustrates that uncertainty in the shock formation radius translates into significant uncertainty in the predicted nucleosynthesis for gravito-acoustic NDWs.

To illustrate the important impact of the reduced electron fraction from the wave contributions, we show in figure~\ref{fig:ye_52} abundance distributions from a wind with $M_{NS}=1.9 M_\odot,~L_\nu=6\times 10^{52},$ and $Y_{e,\textrm{eq}}=0.52$. In the absence of wave effects, the neutrino spectrum used here should preclude any r-processing whatsoever. The wind would undergo an alpha-rich freezeout, leaving only free protons to capture onto seed nuclei. However, with wave effects included, we find similar r-processing regimes to those obtained with neutrino energies tuned to $Y_e=0.48$. In the wave stress regime, with $5\times 10^{-4}\lesssim L_w/L_\nu\lesssim 5\times 10^{-3}$, the change in $Y_e$ is not large enough to make the wind neutron rich, but the faster outflow caused by the wave stress prevents an $\alpha$-rich freezeout from occurring. R-process elements are then synthesized from the free neutrons in the wind, despite the wind being overall proton-rich. This gives rise to the suppressed, actinide-free r-process patterns in figure~\ref{fig:ye_52}. In between the r-processing regimes, we again find a region where the combined entropy and dynamical timescale in the wind favor strong seed formation and thus no r-processing, regardless of $Y_e$ or the presence of an $\alpha$-rich freezeout. At high $L_w$, the wind becomes neutron-rich again, and early wave heating suppresses seed formation and drives the same strong r-processing as in figure~\ref{fig:ext_skynet}.

\subsubsection{Nucleosynthesis in the $L_w$ - $L_\nu$ - $M_\text{NS}$ parameter space}
\begin{figure*}
    \centering
    \includegraphics[scale=1]{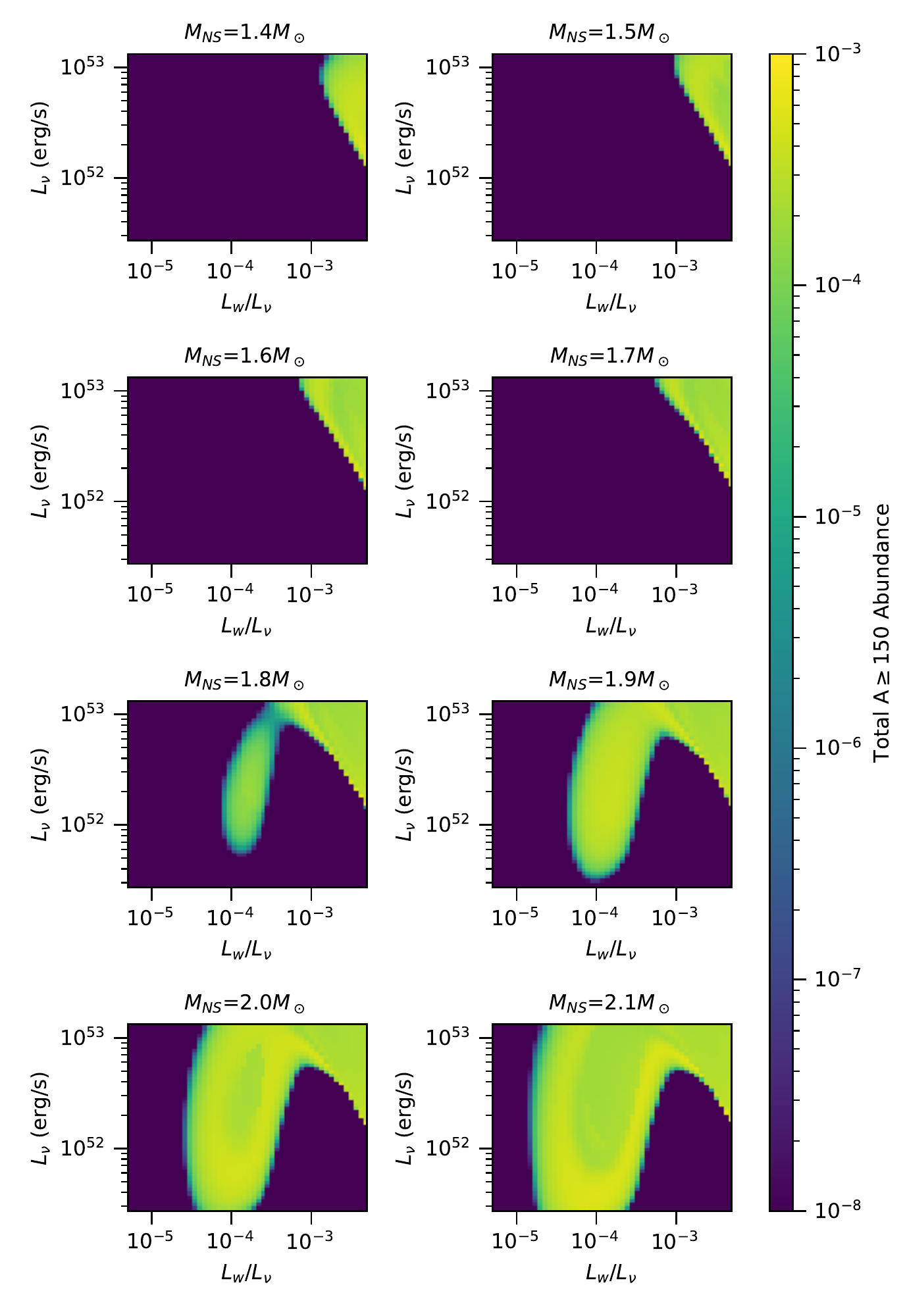}
    \caption{A measure of r-process strength across our parameter space, using a wave frequency of $2 \times 10^3$ rad s$^{-1}$ and $Y_{e, \textrm{eq}} = 0.48$, and using the shock heating prescription from equation (\ref{eq:shock_condition}). Two regimes of r-process production emerge: a region of high wave and neutrino luminosities across all masses, driven by shock heating; and a region of moderate wave and neutrino luminosities driven by the wave stress, which becomes significant at larger neutron star masses. }
    \label{fig:param_space}
\end{figure*}
\begin{figure*}
    \centering
    \includegraphics[scale=1]{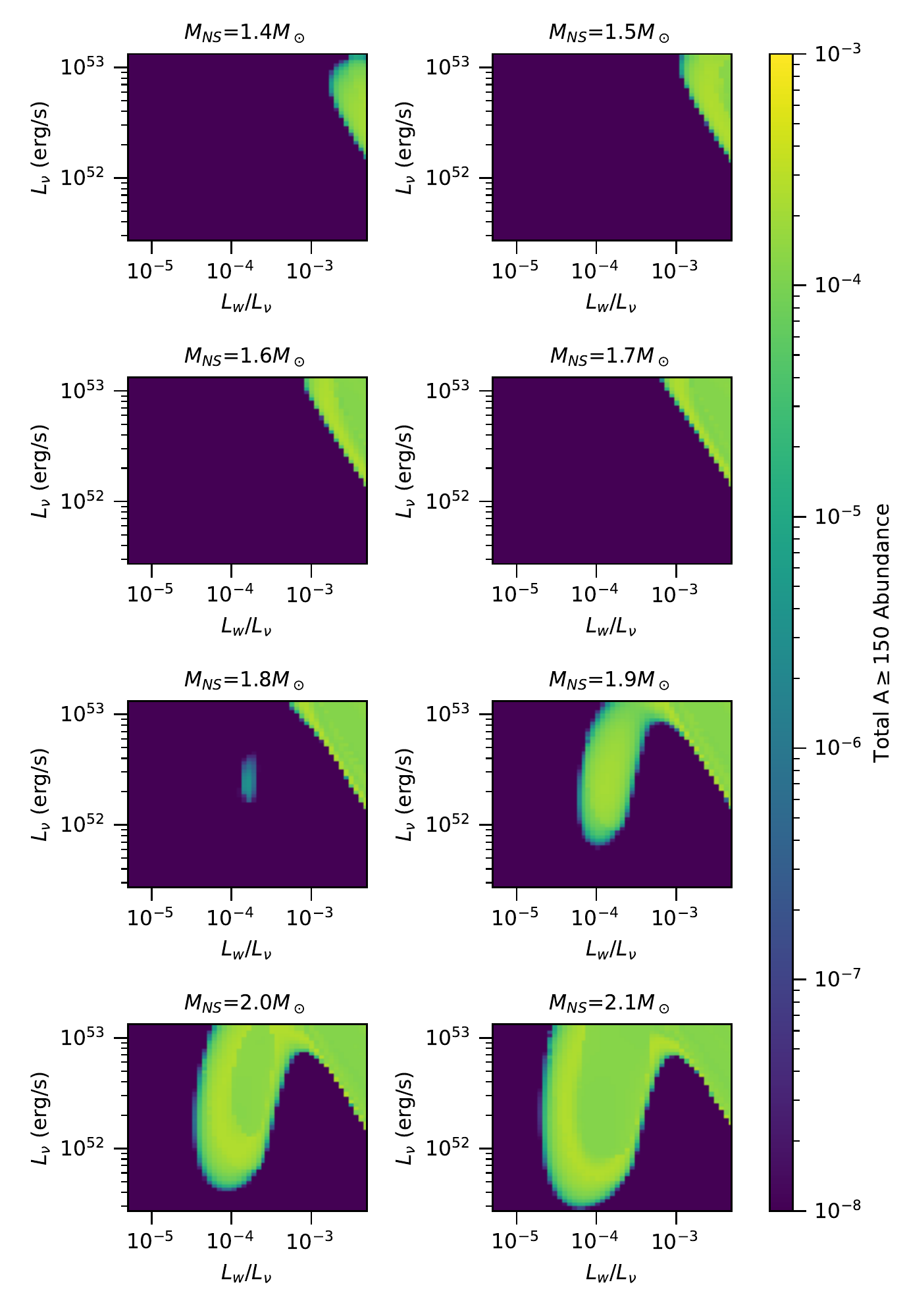}
    \caption{A measure of r-process strength across our parameter space, using identical parameters as figure~\ref{fig:param_space} but with $Y_e$ fixed at 0.48. The same two r-processing regimes emerge, but the wave stress regime is pushed to higher masses and neutrino luminosities by the lowered neutron abundance. }
    \label{fig:param_space_noye}
\end{figure*}
\begin{figure*}
    \centering
    \includegraphics[scale=1]{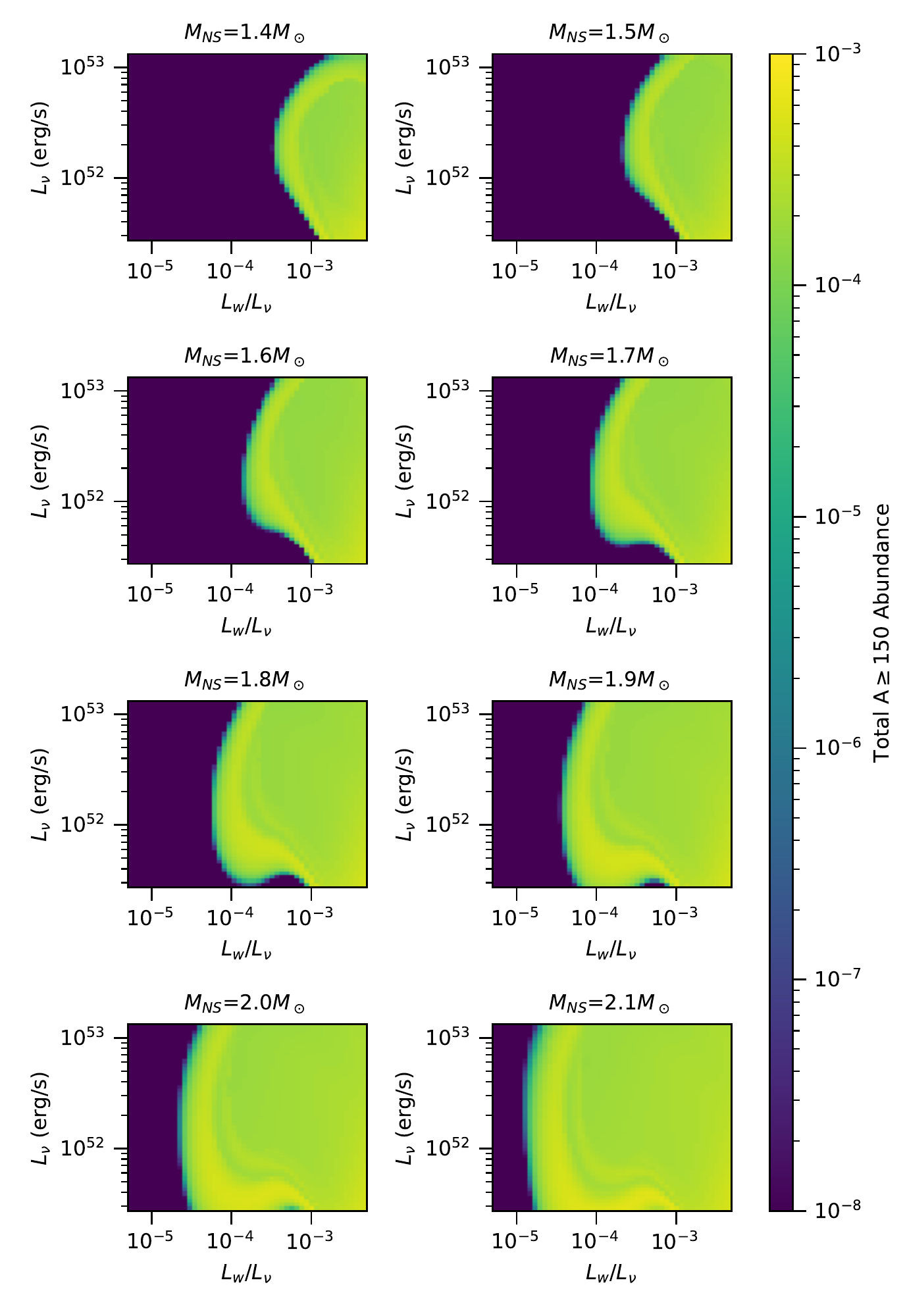}
    \caption{A measure of r-process strength across our parameter space, using a wave frequency of $2\times 10^3$ rad s$^{-1}$ and $Y_{e,\textrm{eq}}=0.48$, and assuming the waves immediately shock and begin to deposit heat into the wind. We see that for higher, but still quite reasonable wave luminosities, r-processing takes place nearly independent of PNS mass and neutrino luminosity. The r-processing parameter space broadens to very low wave luminosities at higher masses as the wave stress becomes significant. }
    \label{fig:instant_skynet_space}
\end{figure*}

In figure~\ref{fig:param_space}, we show the total final abundance of nuclei with mass number $A \geq 150$ as a function of $L_\nu$ and $L_w/L_{\nu}$ for a variety of PNS masses. Here, we have used $Y_{e,\textrm{eq}} = 0.48$, $\omega = 2\times 10^3 \, \textrm{rad s}^{-1}$, and assumed the shock formation radius is given by equation \ref{eq:shock_condition}. We find the abundance of nuclei with $A \geq 150$ to be an effective proxy for the strength of the r-process in the wind (see e.g. figure~\ref{fig:ext_skynet}). Two r-processing regimes appear. For the highest neutrino and wave luminosities, shock heating begins early enough in the wind to drive a strong r-process. This shock heating regime is fairly insensitive to PNS mass but very dependent on wave frequency, which sets how early shock heating can begin in the wind. The second r-processing regime, driven by acceleration due to the wave stress, is strongly dependent on mass but insensitive to wave frequency. We see this regime emerge at a PNS mass of around 1.8$M_\odot$, and grow to dominate the parameter space for the most massive neutron stars. The non-monotonic dependence of the average mass number of the final abundances is also visible here. At higher masses, the wave stress contribution is able to drive strong r-processes even for very low neutrino and wave luminosities, where shock heating begins too late to strongly affect the nucleosynthesis. We have also run similar calculations with $Y_{e,\textrm{eq}} = 0.45$. These show qualitatively similar behavior to the results shown in figure~\ref{fig:param_space}, except that the onset of wave stress-driven r-process nucleosynthesis is shifted to lower PNS mass.

In order to quantify the impact of the reduced electron fraction due to the wave stress contribution, we show in figure~\ref{fig:param_space_noye} the same parameter set as in figure~\ref{fig:param_space}, but with $Y_e$ fixed to a constant value of 0.48. We find that including a self-consistent $Y_e$ evolution results in a noticeable broadening of the region in $L_\nu$-$L_\nu/L_w$ space where the r-process occurs, especially the wave stress-dominated regime at lower $L_w$ and $L_\nu$. This is perhaps to be expected, as the change in $Y_e$ is driven primarily by the wave stress reducing the amount of neutrino heating needed to unbind the wind material. We also observe generally higher yields of r-process material when $Y_e$ evolution is included, due to the higher number of free neutrons available.

Finally, in figure~\ref{fig:instant_skynet_space}, we show the impact of instantaneous shock formation on nucleosynthesis across the entire parameter space (once again with $Y_{e, \textrm{eq}}=0.48$ and $\omega = 2 \times 10^3 \, \textrm{rad s}^{-1}$, and self-consistently evolving $Y_e$). In this case, we find third peak r-process production for nearly all considered neutrino luminosities and PNS masses when $L_w/L_\nu \gtrsim 2 \times 10^{-4}$. Although the acceleration of the wind due to the wave stress plays a role in determining the nucleosynthesis in these models, the impact of the waves is mainly driven by the shock heating that they provide.

\section{Conclusions}

We have investigated the impact of gravito-acoustic waves launched by PNS convection on the dynamics and nucleosynthesis of the neutrino-driven wind. When these waves propagate through the NDW, they impose additional stresses on the wind and also may shock and provide an extra source of heating. Using steady-state, spherically symmetric models for the wind that include the impact of an acoustic wave energy flux, we surveyed the parameter space of the gravito-acoustic wave luminosity and frequency that is expected to be produced by PNS convection. The presence of shock heating in the wind precludes reliance upon the common predictive metric $s^3/\tau_d$, as entropy is no longer nearly constant during seed formation. Therefore, using the results of our hydrodynamic models, we then performed calculations of nucleosynthesis for the marginally neutron-rich compositions that may be encountered in some NDWs.

For $L_w \gtrsim 10^{-5} L_{\nu}$, the waves strongly impact the dynamics of the wind via two mechanisms, acceleration due to wave stresses and entropy production via wave shock heating. Acceleration of the NDW by wave stresses reduces the dynamical timescale, but also reduces the entropy and electron fraction of the wind since a faster wind has less opportunity to undergo neutrino heating. Depending on $L_w/L_{\nu}$, this competition between reduced dynamical timescale and reduced entropy can make conditions more or less favorable for strong r-process nucleosynthesis. 

Similarly to previous work \citep{Suzuki_2005, Metzger_2007}, we find that if the wave energy is deposited (in our case through shock heating) before r-process seed nucleus formation begins, the entropy of the wind at seed formation is substantially increased. This in turn results in an alpha-rich freeze out and more favorable conditions for producing nuclei in the third r-process peak. Here, we found that the exact position of shock formation has a strong impact on the final nucleosynthesis. If wave shock heating begins before a temperature of around $7 \, \textrm{GK}$, the final nucleosynthesis is strongly impacted and even NDWs with modest wave luminosities and fiducial PNS masses can produce a solar-like r-process pattern. If wave shock heating begins below this temperature range, its impacts on nucleosynthesis are muted. For gravito-acoustic waves, the radius of shock formation depends on their frequency, so higher frequency waves are likely to have a larger impact on nucleosynthesis. For higher PNS masses, wave stress contributions can still drive a strong r-process even if shock heating begins too late to affect seed formation.

At high wave luminosities ($L_w\geq 10^{-3}$), the electron fraction can also be reduced by up to almost 10\% as a result of gravito-acoustic wave acceleration of the NDW. This wave-induced reduction in $Y_e$ broadens the regions of $L_w$-$L_\nu$ parameter space over which the r-process occurs and can even cause an r-process to be produced (if conditions are otherwise favorable) in winds with neutrino spectra predicted to result in proton-richness. 

The models we have considered are necessarily approximate, given the substantial uncertainties about the properties of long-term PNS convection and wave propagation in these environments. Nevertheless, they suggest that gravito-acoustic waves may have a significant impact on NDW nucleosynthesis, especially at early times when PNS convection is strongest. As advanced, long-term 3D simulations of core-collapse supernovae become available, our results indicate the importance of resolving and examining the impact of PNS convection on wave excitation and possible NDW dynamics. Of course, the production of the r-process requires the NDW to be at least marginally neutron rich, which recent models suggest may or may not be the case. As a result, we intend to examine the impact of the wave effects discussed here in the context of a proton-rich wind in a subsequent paper.

\section*{Acknowledgements}
BN thanks Edward Brown for helpful discussions during this work. BN acknowledges support from a University Distinguished Fellowship and from the College of Natural Sciences at Michigan State University. We thank Brian Metzger for useful comments. LR thanks Stan Woosley for useful discussions during the early stages of this work. This work was supported in part through computational resources and services provided by the Institute for Cyber-Enabled Research at Michigan State University. This work has been assigned a document release number LA-UR-22-33189.

\section*{Data Availability}
The simulation code and results used in this work are available upon reasonable request to the authors. The SkyNet reaction network used is open-source software publicly available at \url{https://bitbucket.org/jlippuner/skynet}.





\bibliographystyle{mnras}





\bsp	
\label{lastpage}
\end{document}